\documentclass[12pt]{article}
\usepackage{amsmath}
\usepackage{graphicx}
\usepackage{enumerate}
\usepackage{natbib}
\usepackage{url} 

\usepackage{amsmath}
\usepackage{graphicx,psfrag,epsf}
\usepackage{enumerate}
\usepackage{natbib}
\usepackage{url} 
\usepackage{amsthm}
\usepackage{booktabs}
\usepackage{framed}  
\usepackage{caption}
\usepackage{pgfplots}
\usepgfplotslibrary{dateplot}
\usetikzlibrary{snakes}
\usepackage{float}
\usepackage{amsfonts}
\usepackage{multirow, booktabs}
\usepackage{cleveref}
\usepackage{subcaption}
\usepackage[ruled,vlined]{algorithm2e}
\usepackage[T1]{fontenc}
\usepackage[utf8]{inputenc}
\usepackage{authblk}
\usepackage[multiple]{footmisc}
\usepackage{blindtext,titlefoot}
\usepackage{sectsty}
\usepackage{xcolor}
\usepackage{tikz}
\usepackage{amsmath}
\usepackage{graphicx}
\usepackage{comment}
\usepackage{amsfonts}
\usepackage{bbm}
\usepackage{amssymb}
\usepackage{setspace}
\usepackage{natbib}
\usepackage[figuresright]{rotating}
\usepackage{adjustbox}
\usepackage{pifont}
\usetikzlibrary{positioning, shapes.geometric}

\usepackage{tabularx}
\usepackage{ragged2e} 
\newcolumntype{L}{>{\RaggedRight}X} 
\usepackage{lipsum} 

\usepackage[ruled,vlined]{algorithm2e}
\usepackage{amsmath, amssymb}
\usepackage{amsfonts, multirow, epsfig, subfig}
\usepackage{graphicx, pdflscape, verbatim, enumerate, colortbl, setspace}
\usepackage{setspace, color,bm}
\usepackage[normalem]{ulem}
\usepackage{cite}
\usepackage{multirow}
\usepackage{booktabs,array}
\usepackage{url}
\usepackage{bbm}

\newtheorem{prop}{Proposition}
\newtheorem{condition}{Condition}

\newtheorem{lemma}{Lemma}

\newtheorem{theorem}{Theorem}
\newtheorem{example}{Example}
\newtheorem{remark}{Remark}

\makeatletter
\newcommand*{\rom}[1]{\expandafter\@slowromancap\romannumeral #1@}
\makeatother

\begin{document}

\def\spacingset#1{\renewcommand{\baselinestretch}%
{#1}\small\normalsize} \spacingset{1}

\sectionfont{\bfseries\large\sffamily}%
%
\newcommand*\emptycirc[1][1ex]{\tikz\draw (0,0) circle (#1);} 
\newcommand*\halfcirc[1][1ex]{%
  \begin{tikzpicture}
  \draw[fill] (0,0)-- (90:#1) arc (90:270:#1) -- cycle ;
  \draw (0,0) circle (#1);
  \end{tikzpicture}}
\newcommand*\fullcirc[1][1ex]{\tikz\fill (0,0) circle (#1);} 

\subsectionfont{\bfseries\sffamily\normalsize}%
%


\def\spacingset#1{\renewcommand{\baselinestretch}%
{#1}\small\normalsize} \spacingset{1}

\begin{center}
    \Large \bf Bridging the Gap Between Design and Analysis: Randomization Inference and Sensitivity Analysis for Matched Observational Studies with Treatment Doses
\end{center}


\begin{center}
  \large  $\text{Jeffrey Zhang}^{*, 1}$ and $\text{Siyu Heng}^{2}$
\end{center}

\begin{center}
   \large \textit{$^{1}$Department of Statistics and Data Science, University of Pennsylvania}
\end{center}
\begin{center}
   \large \textit{$^{2}$Department of Biostatistics, New York University}
\end{center}

\let\thefootnote\relax\footnotetext{$^{*}$Address for Correspondence: Jeffrey Zhang, Department of Statistics and Data Science, University of Pennsylvania (email: jzhang17@wharton.upenn.edu). }

\bigskip

\begin{abstract}

Matching is a commonly used causal inference study design in observational studies. Through matching on measured confounders between different treatment groups, valid randomization inferences can be conducted under the no unmeasured confounding assumption, and sensitivity analysis can be further performed to assess sensitivity of randomization inference results to potential unmeasured confounding. However, for many common matching designs, there is still a lack of valid downstream randomization inference and sensitivity analysis approaches. Specifically, in matched observational studies with treatment doses (e.g., continuous or ordinal treatments), with the exception of some special cases such as pair matching, there is no existing randomization inference or sensitivity analysis approach for studying analogs of the sample average treatment effect (Neyman-type weak nulls), and no existing valid sensitivity analysis approach for testing the sharp null of no effect for any subject (Fisher's sharp null) when the outcome is non-binary. To fill these gaps, we propose new methods for randomization inference and sensitivity analysis that can work for general matching designs with treatment doses, applicable to general types of outcome variables (e.g., binary, ordinal, or continuous), and cover both Fisher's sharp null and Neyman-type weak nulls. We illustrate our approaches via comprehensive simulation studies and a real-data application.
\end{abstract}

\noindent%

\spacingset{1.73} 

\section{Introduction: Background and Our Main Contributions}

In observational studies, matching is one of the most widely adopted study designs for causal inference. In matched observational studies, subjects with different values of the treatment variable (e.g., treated versus untreated in the binary treatment case, or different treatment doses in the continuous treatment case) are matched for measured confounders so that the treatments as-if randomly assigned within each matched set under the no unmeasured confounding assumption (\citealp{rubin1979using, lu2001matching, rosenbaum_obs, rosenbaum2020design, Baiocchi2010}). After matching, randomization inference can then be conducted as in a randomized experiment under the no unmeasured confounding assumption, and sensitivity analysis can be further conducted to assess sensitivity of randomization inference to unmeasured confounding. 

Originally, matching was mainly designed for handling binary treatments (e.g., \citealp{rubin1979using, rosenbaum_obs}). However, in many practical settings, treatments are non-binary and have treatment doses (e.g., ordinal or continuous treatments). Therefore, throughout the last two decades, many novel matching designs have been proposed for handling observational studies with treatment doses, which tremendously expand the scope of applicability of matching. For example, see \citet{lu2001matching}, \citet{Baiocchi2010}, \citet{rosenbaum2020design}, \citet{greevy2023optimal}, \citet{yu2023risk}, and \citet{zhang2023statistical}, among many others. Although many novel matching designs for handling non-binary treatments have been proposed during the last two decades, the corresponding downstream randomization inference and sensitivity analysis methods are very much underdeveloped. Specifically, after matching with treatment doses, except in some special cases such as pair matching, there is no existing method for randomization inference or sensitivity analysis for studying the sample average treatment effect (i.e., Neyman-type weak nulls) or its analogs, and there is no existing sensitivity analysis method for testing the null of no treatment effect for any subject (i.e., Fisher's sharp null) beyond the non-binary outcome case. Due to a lack of downstream model-free randomization inference techniques and/or corresponding sensitivity analysis methods, previous studies that adopt these emerging matching designs beyond the binary treatment and pair matching settings have to either 1) consider a particular treatment effect model (e.g., linear effect model or its simple variants) when conducting randomization inference or 2) instead turn to some parametric approaches such as running a parametric outcome regression after matching (\citealp{zhang2023statistical}). However, the statistical validity of both of these strategies relies heavily on the correct specification of the treatment effect or the outcome model.

To fill in the aforementioned gaps between design and inference in matched observational studies with treatment doses (e.g., continuous or ordinal treatments), our work the following five main contributions (referred to as \textbf{Contributions 1--5}) to the relevant literature:
\begin{itemize}
    \item For testing the sharp null of no treatment effect (i.e., Fisher's sharp null) in matched observational studies with treatment doses, in Section~\ref{sec: sharp null case}, we propose a new method for sensitivity analysis for unmeasured confounding under the commonly used Rosenbaum bounds framework. This proposed method is the first rigorous sensitivity analysis method that can work for general matching designs with treatment doses (e.g., those beyond pair matching) and general outcome variables (e.g., continuous, ordinal, or binary) (\textbf{Contribution 1}). 

    \item For studying the sample average treatment effect or its analogs (i.e., Neyman-type weak nulls), we first discuss how to generalize the Neyman-type estimands to general matching designs with treatment doses (\textbf{Contribution 2}). We then develop a model-free randomization inference approach for constructing non-parametric estimators and confidence intervals for these new estimands under the no unmeasured confounding assumption (\textbf{Contribution 3}). Our approach allows for incorporating covariate information to improve the efficiency of randomization inference without sacrificing statistical validity. Finally, to further assess sensitivity to unmeasured confounding, we develop new sensitivity analysis methods paired with our randomization inference methods (\textbf{Contribution 4}). These proposed estimands, randomization inference methods, and corresponding sensitivity analysis techniques together form the first randomization inference framework for studying the sample average treatment effect and its analogs for general matching designs with treatment doses (e.g., those beyond the pair matching setting).

    \item In addition to matching with treatment doses, all of the proposed methods in this work also work for stratification with treatment doses (this point will be discussed in detail in Remarks \ref{rem: randomization inference with stratification} and \ref{rem: sensitivity analysis for stratification}). In stratified observational studies with treatment doses, our methods provide 1) the first valid sensitivity analysis method for testing the null effect (i.e., Fisher's sharp null) and 2) the first rigorous randomization inference and corresponding sensitivity analysis frameworks for studying the sample average treatment effect and its analogs (i.e., Neyman-type weak nulls) (\textbf{Contribution 5}). 
\end{itemize}
To summarize, with the proposed methods in this work, practical researchers now have a complete toolkit for randomization inference and corresponding sensitivity analysis to handle matched observational studies with general treatment variables (e.g., binary, ordinal, or continuous), general matching (e.g., pair matching, matching with multiple controls, or full matching) or stratification designs, general outcome variables (e.g., binary, ordinal, or continuous), and for studying either Fisher's sharp null or Neyman-type weak nulls; see Table~\ref{tab: summary of the literature} for a summary. Note that when the treatment variable is binary (i.e., when there are only two treatment doses), although our methods can still provide valid randomization inference and sensitivity analysis results, they may either be more conservative or less computationally efficient than the existing methods specifically designed for the binary treatment case (see Appendix B in the online supplementary materials for details). Therefore, the main contributions of this work still focus on matched or stratified observational studies with treatment doses (e.g., continuous or ordinal treatments) beyond the binary treatment case. 

\setlength{\tabcolsep}{3pt} 
\begin{table}[ht]
\caption{Summary of the scopes of applicability of previous methods and those of our work.}

\centering

\begin{adjustbox}{angle=90}\begin{tabular}{lccccccccc}
\toprule
\multirow{2}{*}{}&\multicolumn{2}{c}{Null Hypothesis}&\multicolumn{3}{c}{Design} & \multicolumn{2}{c}{Treatment} & \multicolumn{2}{c}{Outcome} \\
\cmidrule(rl){2-3} \cmidrule(rl){4-6} \cmidrule(rl){7-8} \cmidrule(rl){9-10} 
 &  Sharp & Weak  & Pair Match & Full Match & Stratification & Binary & Continuous & Binary & Continuous \\
\midrule
Rosenbaum (1987, Biometrika)  & \checkmark  & \ding{55}  & \checkmark  & \checkmark  & \checkmark  &  \checkmark & \ding{55}  & \checkmark   & \ding{55} \\
Rosenbaum (1989, SJS)  & \checkmark  &  \ding{55}  & \checkmark   & \ding{55}   &  \ding{55}  & \checkmark & \checkmark  & \checkmark  & \checkmark \\
Gastwirth et al. (2000, JRSSB)  & \checkmark  & \ding{55} & \checkmark  &  \checkmark &  \checkmark & \checkmark  & \ding{55}  &  \checkmark  & \checkmark \\
Fogarty et al. (2017, JASA) & \checkmark  & \checkmark  & \checkmark  & \ding{55}  & \ding{55}  & \checkmark  & \ding{55} &  \checkmark  & \ding{55} \\
Rosenbaum (2018, AOAS) & \checkmark  & \ding{55}  & \checkmark  & \checkmark  & \checkmark  & \checkmark  & \ding{55}  &  \checkmark  & \checkmark \\
Fogarty (2020, JASA) & \checkmark  & \checkmark &  \checkmark & \ding{55}  & \ding{55}  &  \checkmark &  \ding{55} &  \checkmark  & \checkmark \\
Fogarty (2022, Biometrics) & \checkmark  & \checkmark &  \checkmark & \checkmark  & \ding{55}  &  \checkmark &  \ding{55} &  \checkmark  & \checkmark   \\
Zhang et al. (2024, Biometrika) & \checkmark  & \ding{55}  &  \checkmark & \checkmark  & \checkmark  & \checkmark  & \checkmark  & \checkmark   & \ding{55} \\
Heng and Kang (2024, Preprint) & \checkmark  & \checkmark  & \checkmark  & \ding{55}  & \ding{55}  & \checkmark  & \checkmark  &  \checkmark  & \checkmark \\
Chen et al. (2024, Preprint) &  \checkmark & \checkmark  &  \checkmark &  \ding{55} & \ding{55}  & \checkmark  & \ding{55}  &  \checkmark  & \checkmark \\
\textbf{This Work}  &  \checkmark & \checkmark  &  \checkmark & \checkmark  & \checkmark  & \checkmark  &  \checkmark &  \checkmark  & \checkmark \\
\bottomrule
\end{tabular}\end{adjustbox}
\label{tab: summary of the literature}
\end{table}

Our methods are examined via simulation studies and applied to a matched observational study on the effect of lead exposures on lumbar bone mineral density (BMD) among females aged 20--39. All the proposed methods in this work have been incorporated into publicly available $\texttt{R}$ package \texttt{doseSens}.

\section{Review} 

\subsection{Randomization Inference for Matched Observational Studies with Treatment Doses}

We first review the randomization inference framework for matched observational studies with continuous treatments, following the notations considered in existing work \citep{rosenbaum1989sensitivity, zhang2024sensitivity}. After matching, suppose that there are $I$ matched sets. In matched set $i$ ($i=1,\dots, I$), there are $n_i$ subjects, giving a total of $N=\sum_{i=1}^I n_i$ subjects. For each subject $j$ in matched set $i$ (henceforth called subject $ij$), we let $\mathbf{x}_{ij}$ denote the vector of measured confounders, $Z_{ij}$ the observed treatment dose, and $R_{ij}$ the observed outcome. Let $\mathbf{Z}=(Z_{11},\dots, Z_{In_{I}})$ and $\mathbf{R}=(R_{11},\dots, R_{In_{I}})$ denote the treatment dose vector and the observed outcome vector, respectively. In matching with treatment doses, each matched set $i$ has $n_i$ observed treatment doses $Z_{i1}, \dots, Z_{in_{i}}$, either all distinct (e.g., in the continuous treatment case) or may have ties (e.g., in the ordinal treatment case). Let $\widetilde{Z}$ represent the set of all possible treatment doses in the study. Following the potential outcomes framework \citep{neyman1923application, rubin1974estimating, rosenbaum1989sensitivity}, we let $r_{ij}(z)$ denote the potential outcome of subject $ij$ under the treatment dose $z\in \widetilde{Z}$. Then, the observed outcome $R_{ij}=r_{ij}(z)$ when $Z_{ij}=z$. Under the randomization inference framework, all the potential outcomes $r_{ij}(z)$ are fixed values, and the only source of randomness that enters into inference is the distribution of treatment dose assignments \citep{rosenbaum_obs, gastwirth1998dual, zhang2023statistical}. We let $\mathcal{F}_{0}=\{\mathbf{x}_{ij},r_{ij}(z),i=1,...,I,j=1,\dots, n_{i}, z \in \widetilde{Z}\}$ denote the collection of the information of all measured confounders and potential outcomes. In each matched set $i$, given the observed treatment doses $\mathbf{z}_{i}=(z_{i1}, \dots, z_{in_{i}})$, corresponding to $n_i!$ permutations of $\mathbf{z}_{i}$, there are $n_{i}!$ possible realizations of the treatment dose vector $\mathbf{Z}_{i}=(Z_{i1}, \dots, Z_{in_{i}})$ (here, for notations simplicity, when there are ties among treatment doses in $\mathbf{z}_{i}$ and different permutations of $\mathbf{z}_{i}$ may correspond to the same vector value, we still conceptually treat them as different permutations of $\mathbf{z}_{i}$; this is only for making the notations universal and will not affect any conclusions in this work). Let $S_{n_i}$ denote the set of all $n_{i}!$ permutations of the index set $\{i1, \dots, in_{i}\}$ and $\mathcal{Z}_{i}=\{\mathbf{z}_{i\pi_{i}} \mid \pi_{i} \in S_{n_i}\}$ the set of all $n_{i}!$ permutations of $\mathbf{z}_{i}$. For example, if a matched set has three observed treatment doses 0.2, 0.5, and 0.7, then there are $3!$ possible assignments of these three doses to the three subjects in randomization (permutation) inference: $\mathcal{Z}_{i}=\{(0.2, 0.5, 0.7), (0.2, 0.7, 0.5), (0.5, 0.2, 0.7), (0.5, 0.7, 0.2), (0.7, 0.2, 0.5), (0.7, 0.5, 0.2)\}$. To simplify the notations, we let $z_{i(k)}$ denote the $k$-th order statistic of the treatment doses in matched set $i$ (i.e., we have $z_{i(1)}\leq \dots \leq z_{i(k)} \leq \dots \leq z_{i(n_{i})} $) and $r_{ij}^{(k)} = r_{ij}(z_{i(k)})$ the potential outcome of subject $ij$ under dose $z_{i(k)}$. Let $\mathcal{Z}=\mathcal{Z}_{1}\times \mathcal{Z}_{2} \times \dots \times \mathcal{Z}_{I}$ denote all possible dose assignments in randomization inference for the whole matched dataset (e.g., when there are no ties among doses, the cardinality of $\mathcal{Z}$ equals $|\mathcal{Z}|=n_{1}!\dots n_{I}!$). 

Under the no unmeasured confounding assumption and matching on the measured confounders (i.e., $\mathbf{x}_{ij}=\mathbf{x}_{ij^{\prime}}$ or $\mathbf{x}_{ij}\approx \mathbf{x}_{ij^{\prime}}$ for all $j,j^{\prime} \in \{1,\dots, n_i\}$), the treatment dose assignments are (or are nearly) completely random within each matched set $i$:
\begin{equation*}
\text{$p_{i\pi_{i}}:=P(\mathbf{Z}_i=\mathbf{z}_{i\pi_{i}}|\mathcal{F}_{0},\mathcal{Z})=\frac{1}{n_{i}!}$ for all $\pi_{i} \in S_{n_i}$ (i.e., for all $\mathbf{z}_{i\pi_{i}} \in \mathcal{Z}_{i}$).}
\end{equation*}
Assuming independence of treatment dose assignments across matched sets, we have 
\begin{equation}\label{eqn: random assignment assumption}
    \text{$P(\mathbf{Z}=\mathbf{z} |\mathcal{F}_{0},\mathcal{Z})=\frac{1}{n_{1}!\dots n_{I}!}$ for all $\mathbf{z} \in \mathcal{Z}$.}
\end{equation}
 Then, researchers can conduct randomization inference under the above random treatment dose assignment assumption (\ref{eqn: random assignment assumption}) after matching. For example, for testing the null hypothesis of no treatment effect for all subjects (i.e., Fisher's sharp null) 
 \begin{equation*}
     \text{$H_F: r_{ij}(z)=r_{ij}(z^{\prime})$ for all $i, j,$ and $z, z^{\prime} \in \mathcal{Z}_{i}$,}
 \end{equation*}
 given the test statistics $T(\mathbf{Z}, \mathbf{R})$ and its observed value $t$, the corresponding permutation $p$-value is $P(T \geq t|\mathcal{F}_{0},\mathcal{Z})=|\{\mathbf{z} \in \mathcal{Z}: T(\mathbf{Z}=\mathbf{z}, \mathbf{R}) \geq k\}|/(n_{1}!\dots n_{I}!)$ (\citealp{rosenbaum1989sensitivity, gastwirth1998dual, zhang2023statistical}), which can be approximated by the Monte-Carlo method when sample size $N$ is large.

 In addition to testing Fisher's sharp null $H_{F}$, researchers often also study the sample average treatment effect (i.e., Neyman-type weak nulls). For example, in pair matching (i.e., when $n_{i}=2$ for all $i$) with treatment doses, researchers often consider hypothesis testing and randomization inference for the effect ratio \citep{Baiocchi2010, rosenbaum2020design, Zhang2022BridgingRatio}, which can be regarded as a generalization of the classic SATE (i.e., Neyman's weak null problem) from the binary treatment setting to the general treatment settings:
\begin{equation}\label{eqn: effect ratio}
   \text{Effect Ratio } \lambda = \frac{\sum_{i=1}^I\sum_{j=1}^2 (r_{ij}^{(2)}-r_{ij}^{(1)})}{\sum_{i=1}^I \sum_{j=1}^2 (z_{i(2)}-z_{i(1)})} \text{ (recall that $r_{ij}^{(k)}=r_{ij}(z_{i(k)})$)}.
\end{equation}
That is, the effect ratio $\lambda$ is the ratio between the total difference in potential outcomes under the paired treatment doses and the total difference in paired treatment doses. When the treatment is binary (i.e., when there are only two treatment doses), the SATE $ \lambda$ reduces to the classic Neyman's ATE in finite-sample. Researchers then can adopt a generalized difference-in-means estimator and the corresponding Neyman-type variance estimator to conduct inference for $\lambda$ \citep{Baiocchi2010, Zhang2022BridgingRatio}. 

However, beyond the pair matching case (in which we have a flexible causal estimand -- effect ratio), all the existing causal estimands in randomization inference for matching with treatment doses are constant effects or their related variants (\citealp{zhang2023statistical, zhang2024sensitivity}). That is, there is still a lack of sensible estimands that can serve as analogs of the sample average treatment effect (i.e., Neyman-type weak nulls), as well as their corresponding randomization inference methods. We will fill this important gap in Section~\ref{sec: weak null case}.

 \begin{remark}\label{rem: randomization inference with stratification}
In some stratified observational studies, researchers also assume the randomization assumption (\ref{eqn: random assignment assumption}) after stratifying on measured confounders. In this setting, randomization inference can be similarly conducted assuming no unmeasured confounding (\citealp{rosenbaum2018sensitivity}). Therefore, our proposed randomization inference approaches in Section~\ref{sec: weak null case} can also be applicable in stratified observational studies.  
 \end{remark}

\subsection{Sensitivity Analysis for Matched Observational Studies with Treatment Doses}

Although matching can effectively adjust for measured confounders, unmeasured confounders may exist in many cases and may bias the randomization distribution (\ref{eqn: random assignment assumption}) of treatment dose assignments. Therefore, after conducting randomization inference based on the randomization assumption (\ref{eqn: random assignment assumption}), researchers often conduct a sensitivity analysis to assess the sensitivity of randomization inference results to potential violations of the randomization assumption (\ref{eqn: random assignment assumption}) due to unmeasured confounding. In matched observational studies with treatment doses, a commonly used sensitivity analysis model is the following Rosenbaum treatment dose model \citep{rosenbaum1989sensitivity, gastwirth1998dual} that models the generalized propensity score of treatment dose $Z$ given the measured confounder $\mathbf{x}$ and a hypothetical unmeasured confounder $u$:
\begin{equation}\label{eqn: Rosenbaum treatment dose model}
    P(Z=z|\mathbf{x},u)=\zeta(\mathbf{x},u) \eta(z,\mathbf{x})\exp(\gamma z u)\propto \eta(z,\mathbf{x})\exp(\gamma z u),
\end{equation} 
where the generalized propensity score $P(Z=z|\mathbf{x},u)$ represents the conditional distribution (if $Z$ is discrete) or conditional density (if $Z$ is continuous) of $Z$, the $\zeta(\mathbf{x},u)=1/\int \eta(z,\mathbf{x})\exp(\gamma z u)dz$ is a normalizing constant, the $\eta(z,\mathbf{x})$ is an arbitrary nuisance function, the $\gamma$ is the sensitivity parameter, and the hypothetical unmeasured confounder $u \in [0,1]$ is normalized to make the sensitivity parameter $\gamma$ more interpretable. The Rosenbaum treatment dose model (\ref{eqn: Rosenbaum treatment dose model}) includes many existing common treatment models as special cases. For example, under the no unmeasured confounding assumption (i.e., when $\gamma=0$), model (\ref{eqn: Rosenbaum treatment dose model}) contains the most general model for the generalized propensity score. When the treatment is binary, model (\ref{eqn: Rosenbaum treatment dose model}) reduces to the classic Rosenbaum propensity score model (\ref{eqn: Rosenbaum treatment dose model}). When the treatment is ordinal, model (\ref{eqn: Rosenbaum treatment dose model}) contains Heberman's log-linear models (\citealp{haberman1974log}), which are commonly considered in matched studies (\citealp{rosenbaum1989sensitivity, gastwirth1998dual}). For continuous treatments, model (\ref{eqn: Rosenbaum treatment dose model}) contains the generalized partially linear model: $Z=h(\mathbf{x})+\gamma\cdot u + \epsilon$ for some arbitrary and possibly unknown function $h$ and normal error $\epsilon$. See \citet{rosenbaum1989sensitivity} for detailed discussions. 

For subject $ij$, let $u_{ij}\in [0,1]$ denote its normalized unmeasured confounder. Let $\mathbf{u}=(u_{11},\dots, u_{In_{I}})\in [0,1]^{N}$ denote the collection of unmeasured confounders. We let $\mathcal{F}=\{\mathbf{x}_{ij}, u_{ij}, r_{ij}(z),i=1,\dots,I, j=1,\dots, n_{i}, z \in \widetilde{Z}\}$. Under the Rosenbaum treatment dose model (\ref{eqn: Rosenbaum treatment dose model}), it is easy to show that the treatment dose assignment probability within each matched set $i$, for any treatment doses assignment $\mathbf{z}_{i\pi_{i}} \in \mathcal{Z}_{i}=\{\mathbf{z}_{i\pi_{i}} \mid \pi_{i} \in S_{n_i}\}$, is
\begin{equation}
\label{eqn: dose assignment in matched set i after matching}p_{i\pi_{i}}:=P(\mathbf{Z}_i=\mathbf{z}_{i\pi_{i}}|\mathcal{F},\mathcal{Z}_{i})=\frac{\exp\{\gamma(\mathbf{z}_{i\pi_{i}}\mathbf{u}_i^{T})\}}{\sum_{\mathbf{z}_{i\pi_{i}}^{\prime} \in \mathcal{Z}_i}\exp\{\gamma(\mathbf{z}_{\pi_{i}}^{\prime}\mathbf{u}_i^{T})\}}, 
\end{equation}
where $\mathbf{u}_{i}=(u_{i1}, \dots, u_{in_{i}})\in [0,1]^{n_{i}}$. Assuming independence across matched sets, we have 
\begin{equation}\label{eqn: dose assignment after matching}
    P(\mathbf{Z}=\mathbf{z}|\mathcal{F},\mathcal{Z})=\prod_{i=1}^{I} \frac{\exp\{\gamma(\mathbf{z}_{i}\mathbf{u}_i^{T})\}}{\sum_{\mathbf{z}_{i\pi_{i}}^{\prime} \in \mathcal{Z}_i}\exp\{\gamma(\mathbf{z}_{i\pi_{i}}^{\prime}\mathbf{u}_i^{T})\}} \quad \text{for any $\mathbf{z}=(\mathbf{z}_{1}, \dots, \mathbf{z}_{I})\in \mathcal{Z}$}.
\end{equation}
For example, in pair matching, model (\ref{eqn: dose assignment after matching}) implies that the odds ratio of receiving the higher dose versus that of the lower dose in matched set $i$ is bounded above by $\exp\{|\gamma(Z_{i1}-Z_{i2})|\}$ and below by $\exp\{-|\gamma(Z_{i1}-Z_{i2})|\}$. When the treatment is binary, model (\ref{eqn: dose assignment after matching}) reduces to the classic Rosenbaum biased treatment assignment model \citep{rosenbaum_obs}. 

In sensitivity analysis for matched observational studies, there are two central tasks: Under model (\ref{eqn: Rosenbaum treatment dose model}) (or, equivalently, model (\ref{eqn: dose assignment after matching})), for each prespecified value of sensitivity parameter $\gamma$, among all possible allocations of unmeasured confounders $\mathbf{u}$, researchers aim to solve (i) the worst-case (largest) $p$-value under Fisher's sharp null of no treatment effect $H_{F}$ and (ii) the worst-case (widest) confidence interval for analogs of the SATE (i.e., Neyman-type weak nulls) (e.g., the effect ratio in the pair matching case).

However, the solutions to the aforementioned two tasks are still unknown in matched observational studies with treatment doses, except in some special cases such as pair matching. Specifically, except for pair matching, there is still a lack of valid methods for solving the worst-case $p$-value for testing Fisher's sharp null of no effect $H_{F}$ or the worst-case (widest) confidence interval for analogs of the SATE (i.e., Neyman-type weak nulls). Moreover, beyond the pair matching case, sensible estimands that can be regarded as analogs of the SATE (i.e., Neyman-type weak nulls) have not been proposed, let alone corresponding randomization inference and sensitivity analysis methods. We will fill all these gaps in both the sharp null case (in Section \ref{sec: sharp null case}) and the weak null case (in Section \ref{sec: weak null case}).

\begin{remark}\label{rem: sensitivity analysis for stratification}
  In addition to matched observational studies, the Rosenbaum sensitivity analysis model (\ref{eqn: Rosenbaum treatment dose model}) has also been adopted in stratified observational studies (\citealp{rosenbaum2018sensitivity}). Therefore, all the proposed sensitivity analysis methods in Sections \ref{sec: sharp null case} and \ref{sec: weak null case} can also be directly applied to stratified observational studies. 
\end{remark}

\section{Universally Applicable Sensitivity Analysis for Testing Fisher's Sharp Null of No Effect in Matched Observational Studies with Treatment Doses }\label{sec: sharp null case}

In this section, we propose the first valid sensitivity analysis method for testing Fisher's sharp null $H_F$ that can be applicable in general matching designs with treatment doses and work for general outcome variables (e.g., binary, ordinal, or continuous outcomes). Specifically, given a test statistic $T(\mathbf{Z}, \mathbf{R})$ and its observed value $t$, and for a prespecified sensitivity parameter $\Gamma=\exp(\gamma)$, we provide an asymptotic upper bound for the worst-case $p$-value under Fisher's sharp null $H_F$: 
\begin{equation}\label{eqn: worst-case p-value}
    \max_{\mathbf{u}\in [0,1]^{N} }P(T\geq t \mid H_{F}, \mathcal{F}, \mathcal{Z})= \max_{\mathbf{u}\in [0,1]^{N} }\sum_{\mathbf{z}\in \mathcal{Z}} \Big[ \mathbbm{1}\{T(\mathbf{Z}=\mathbf{z}, {R})\geq t\}\times P(\mathbf{Z}=\mathbf{z}|\mathcal{F},\mathcal{Z}) \Big].
\end{equation}
Previous work (\citealp{rosenbaum1990sensitivity, Gastwirth2000AsymptoticAnalysis}) has shown that, in the binary treatment case, the solution to the worst-case $p$-value (\ref{eqn: worst-case p-value}) must lie in $\{0,1\}^N$ (i.e., at the boundary of the search region of $\mathbf{u}\in [0,1]^{N}$) for a large class of test statistics $T$. However, this nice property no longer holds when treatment is non-binary. Specifically, \citet{zhang2024sensitivity} constructed a counterexample where the solution to the worst-case $p$-value (\ref{eqn: worst-case p-value}) does not lie at the boundary $\{0,1\}^N$ of the search region $[0,1]^N$. Furthermore, \citet{zhang2024sensitivity} showed that solving the worst-case $p$-value (\ref{eqn: worst-case p-value}) in the continuous treatment case is asymptotically equivalent to solving a signomial program, known to be computationally infeasible in practice. Therefore, instead of aiming for an exact solution to the worst-case $p$-value (\ref{eqn: worst-case p-value}), we turn to construct an asymptotic upper bound for (\ref{eqn: worst-case p-value}) to facilitate an asymptotically valid sensitivity analysis for testing $H_{F}$. 

Specifically, we will consider a general class (denoted as $\mathcal{T}$) of test statistics $T(\mathbf{Z}, \mathbf{R}) = I^{-1}\sum_{i=1}^I T_i$, where $T_i = \sum_{j=1}^{n_i} q_{i1}(Z_{ij}) q_{i2}(R_{ij})$ with each $q_{i1}, q_{i2}$ being some prespecified functions. The class $\mathcal{T}$ includes a wide range of commonly used test statistics. For example, if we take $q_{i1}$ and $q_{i2}$ to be the identity function, we get the permutational $t$-test. If we set $q_{i1}$ to be the identity function and $q_{i2}$ to be the outcome rank function (the function that maps each outcome to its rank among all outcomes, i.e., $q_{i2}(r) = \sum_{i=1}^I \sum_{j=1}^{n_i} \mathbbm{1}\{r \geq R_{ij}\}$), we get the classic Wilcoxon rank sum test. If we let $q_{i1}$ be the dose rank function and $q_{i2}$ the outcome rank function, we get the double-rank test (\citealp{rosenbaum1989sensitivity, zhang2023statistical}). We then let $t_{i\pi_{i} } = \sum_{j=1}^{n_i}q_{i1}(z_{i(\pi_{i}(j))})q_{i2}(r_{ij}^{\pi_{i}(j)})$ denote the value $T_i$ would take under the dose assignment $\mathbf{z}_{\pi_{i}}\in \mathcal{Z}_{i}$ within matched set $i$ (or equivalently, under permutation $\pi_{i}\in S_{n_{i}}$), in which all of the $t_{i\pi_{i}}$ are known and fixed under Fisher's sharp null $H_{F}$. Therefore, we have $E[T_i \mid \mathcal{F},\mathcal{Z}] = \sum_{\pi_{i}\in S_{n_{i}} } p_{i\pi_{i}}t_{i\pi_{i}}$. When unmeasured confounding exists, the post-matching treatment dose assignment probabilities $p_{i\pi_{i} }$ will depart from the uniform distribution $1/n_{i}$ and are unknown. However, under the Rosenbaum sensitivity model (\ref{eqn: dose assignment in matched set i after matching}), we can derive the following informative constraints on $p_{i\pi_{i}}$. 

\begin{lemma}
For any post-matching dose assignment probabilities $p_{i\pi_{i}}$ and $p_{i\pi_{i}^{\prime}}$, under the Rosenbaum sensitivity model (\ref{eqn: dose assignment in matched set i after matching}) with the prespecified sensitivity parameter $\gamma>0$, we have
\begin{equation}\label{eqn: linear constraints}
\frac{p_{i\pi_{i}}}{p_{i\pi_{i}^{\prime}}} \in \Big[\exp\Big\{\gamma\Big(\sum_{j: z_{i\pi_{i}(j)} 
< z_{i\pi'_{i}(j)}}(z_{i\pi_{i}(j)}-z_{i\pi'_{i}(j)})\Big)\Big\}, \exp\Big\{\gamma\Big(\sum_{j: z_{i\pi_{i}(j)} 
> z_{i\pi'_{i}(j)}}(z_{i\pi_{i}(j)}-z_{i\pi'_{i}(j)})\Big)\Big\}\Big].
\end{equation}
\end{lemma}
For each matched set $i$ and the prespecified sensitivity parameter $\gamma$ (or, equivalently, $\Gamma=\exp(\gamma)$), we can solve the following linear program to find an upper bound for $E[T_i \mid \mathcal{F},\mathcal{Z}]$:
\begin{equation}\label{eqn: linear program}
\begin{aligned}
& \underset{p_{i\pi_{i}}}{\text{maximize}}
& & \sum_{\pi_{i}\in S_{n_{i}}}p_{i\pi_{i}}t_{i\pi_{i}}\\
& \text{subject to} & &\sum_{\pi_{i}\in S_{n_{i}}} p_{i\pi_{i} } = 1 \text{ and the bounding constraints (\ref{eqn: linear constraints}).}
\end{aligned}
\end{equation}
Let $\mu_i^{*}$ denote the optimal value of the linear program (\ref{eqn: linear program}). Denote $V_F = T- I^{-1}\sum_{i=1}^I \mu_i^{*}= I^{-1}\sum_{i=1}^I V_{F,i}$, in which $V_{F, i} = T_i - \mu_i^{*}$. Then, we have $E_{\mathbf{u}} (V_F \mid H_{F}, \mathcal{F}, \mathcal{Z}) \leq 0$ for any $\mathbf{u}$. 

The next step is to construct a valid variance estimator for $V_{F}$. We propose to utilize a covariate-adjusted variance estimate for $V_{F}$, inspired by the arguments in \citet{Fogarty2018OnExperiments} in the binary treatment case. Such a covariate-adjusted variance estimator can incorporate covariate information to improve efficiency and, meanwhile, still ensure statistical validity. Specifically, let $Q$ be an arbitrary $I \times L$ matrix, with $L < I$. Such a matrix $Q$ allows researchers to incorporate covariate information $\mathbf{x}_{ij}$ to improve efficiency. For example, we can set $Q=(\mathbf{1}_{I\times 1}, \overline{\mathbf{x}}_{1}, \dots, \overline{\mathbf{x}}_{K})$, in which we assume the dimension of $\mathbf{x}_{ij}=(x_{ij1}, \dots, x_{ijK})$ (denoted as $K$) is less than $I-1$ and define $\overline{\mathbf{x}}_{k}=({n_1}^{-1}\sum_{j=1}^{n_{1}}x_{1jk}, \ldots, {n_I}^{-1}\sum_{j=1}^{n_{I}}x_{Ijk})^T$ as the length $I$ vector containing the mean value of the $k$-th covariate in matched set $i$ (for each $k=1,\dots, K$). Next, let $H_Q = Q(Q^TQ)^{-1}Q^T$ be the the hat matrix for $Q$, and $h_{Q_{ij}}$ the $(i,j)$ element of $H_Q$. Let $y_{F,i} = V_{F,i}/ \sqrt{1-h_{Q_{ii}}}$, $\mathbf{y}_{F}=(y_{F,1}, \dots, y_{F,I})$, $\mathcal{I}$ an $I\times I$ identity matrix, and $W$ a $I\times I$ diagonal matrix with the $i$th diagonal entry $w_{i}$ equalling $I n_i / N$. Finally, we define the variance estimator $S_F^2(Q) = I^{-2}\mathbf{y}_{F}W(\mathcal{I}-H_Q)W\mathbf{y}^{T}_{F}$. By adjusting the proof of Proposition 1 in \citet{Fogarty2018OnExperiments} to the setting of the non-binary treatment case and the observational study with treatment dose, we can show that $S_F^2(Q)$ is a valid estimator for $\text{Var}_{\mathbf{u}}(V_F \mid 
    \mathcal{F}, \mathcal{Z})$ for any $\mathbf{u}$.
\begin{lemma}
\label{lem: conservative variance}
Assuming independence of treatment dose assignments across matched sets. For any $I\times L$ matrix $Q$ ($L<I$), we have $E_{\mathbf{u}}[S_F^2(Q) \mid \mathcal{F}, \mathcal{Z}] \geq \text{Var}_{\mathbf{u}}(V_F \mid 
    \mathcal{F}, \mathcal{Z})$ for any $\mathbf{u}$.
\end{lemma}

After constructing a test statistic $V_F$ with non-positive expectation under $H_F$, as well as a valid (i.e., conservative) variance estimator $S_F^2(Q)$ for $V_{F}$, we define $\overline{p}_{F}=1-\Phi(V_F / S_F^2(Q))$ where $\Phi$ is the distribution function of $N(0,1)$. We then show that $\overline{p}_{F}$ is an asymptotic upper bound for the worst-case $p$-value (\ref{eqn: worst-case p-value}) under $H_{F}$, which is equivalent to the following theorem: 
\begin{theorem}
Under the Rosenbaum sensitivity model (\ref{eqn: dose assignment after matching}) with a prespecified sensitivity parameter $\Gamma=\exp(\gamma)$, as well as some mild regularity conditions outlined in the supplementary material, we have $\overline{p}_{F}$ is an asymptotic upper bound for the worst-case $p$-value (\ref{eqn: worst-case p-value}), or equivalently, we have $\lim_{I \to \infty} P_{\mathbf{u}}[\overline{p}_{F}\leq \alpha \mid H_{F}, \mathcal{F}, \mathcal{Z}] \leq \alpha$ for any $\alpha\in (0,1)$ and $\mathbf{u}$.
\end{theorem}
Therefore, we can use $\overline{p}_{F}$ to conduct a valid sensitivity analysis for testing $H_{F}$ in matching with treatment doses, which will be illustrated in simulations studies in Section~\ref{sec: sharp null case}. 


\section{Randomization Inference and Sensitivity Analysis for Neyman-Type Weak Nulls in Matched Observational Studies with Treatment Doses}\label{sec: weak null case}

 \subsection{Generalizing Neyman-type weak nulls and estimands to matched observational studies with treatment doses}
 
 In matched observational studies with binary treatment, the sample average treatment effect (SATE) $N^{-1}\sum_{i=1}^{I}\sum_{j=1}^{n_{i}}\{r_{ij}(1)-r_{ij}(0)\}$ is arguably the most commonly used estimand for measuring some overall treatment effect among the matched subjects, and the null hypothesis SATE $=\lambda_{0}$ for some prespecified number is also referred to as Neyman's weak null (\citealp{neyman1923application, rosenbaum_obs}). In matched observational studies with treatment doses and a pair-matching design, a commonly used estimand is the effect ratio (\ref{eqn: effect ratio}), which can be viewed as an analog of the SATE when treatment doses exist. However, beyond the pair-matching case, when treatment doses exist, there is still a lack of a sensible estimands that follow the spirit of the SATE or Neyman-type weak nulls. Because of this, outside of the pair matching case, the existing randomization inference framework for matching with treatment doses can only test Fisher's sharp null or specific parametric treatment effects models that impute all potential outcomes, such as a linear proportional treatment effect model (\citealp{zhang2023statistical}).
 
 To fill this gap, we propose a broad class of estimands that includes the SATE and effect ratio as special cases while accommodating general treatment variables (e.g., binary, ordinal, and continuous) and general matching designs (including those beyond pair matching). The estimands we study take the following form:
 \begin{equation}
 \label{eq: weak_null_estimand}
     \theta=\sum_{i=1}^I \frac{n_i}{N} \theta_i =\frac{1}{N}\sum_{i=1}^I  \sum_{j=1}^{n_i} \sum_{k=1}^{n_i} f_i^{(k)}(z_{i(k)},r_{ij}^{(k)}),
 \end{equation}
where $\theta_i = n_{i}^{-1}\sum_{j=1}^{n_i} \sum_{k=1}^{n_i} f_i^{(k)}(z_{i(k)},r_{ij}^{(k)})$ represents the contribution to $\theta$ from matched set $i$ (which will be weighted by the sample size proportion $n_{i}/N$), and $f_i^{(k)}(z_{i(k)},r_{ij}^{(k)})$ represents some prespecified function of which the form may vary across different $i=1,\dots, I$ and $k=1,\dots, n_{i}$. Intuitively, in each matched set $i$, there are $n_i$ subjects who each have $n_i$ terms of the form $ c_i(j,k) := f_i^{(k)}(z_{i(k)},r_{ij}^{(k)})$ corresponding to the $n_i$ potential outcomes under $r_{ij}^{(1)}$, \dots, $r_{ij}^{(n_{i})}$, among which only $r_{ij}(Z_{ij}=z_{ij})$ is observed and the other $n_i-1$ potential outcomes are counterfactuals and unobserved. The estimand $\theta$ averages these observed and unobserved terms for each subject, then averages over the subjects in each matched set and, finally, averages over the matched sets. Below, we unpack this general estimand and provide several examples. These example estimands can be regarded as generalizations of existing commonly used estimands to the general treatment variables and general matching designs.

\begin{example}[Sample Average Treatment Effect (SATE)]

 The general estimand $\theta$ defined in (\ref{eq: weak_null_estimand}) covers the SATE when the treatment is binary (i.e., when there are only two doses). Suppose matched set $i$ has $\sum_{j=1}^{n_{i}} Z_{ij} = m_i$ (i.e. $m_i$ treated units) where $0 < m_i < n_i$. Then the $n_i - m_i$ smallest doses are 0, and the $m_i$ largest doses are 1. In the general form (\ref{eq: weak_null_estimand}), setting $f_i^{(k)}(z,r) = -(n_i- m_i)^{-1} \times r$ if $k \leq n_i -m_i$ and $f_i^{(k)}(z,r) = m_i^{-1} \times r$ otherwise gives
 \begin{equation*}
     \theta = \frac{1}{N}\sum_{i=1}^I\sum_{j=1}^{n_i} \Big\{\sum_{k=n_i-m_i+1}^{n_i}\frac{r_{ij}^{(k)}}{m_i}-\sum_{k=1}^{n_i-m_i}\frac{r_{ij}^{(k)}}{n_i-m_i}\Big\}= \frac{1}{N}\sum_{i=1}^I\sum_{j=1}^{n_i} \{r_{ij}(1) - r_{ij}(0)\},
 \end{equation*}
 which is exactly the definition of the SATE.
\end{example}

\begin{example}[Effect Ratio]
In the pair-matching case (i.e., $n_{i}=2$ for all $i$), setting $ f_i^{(2)}(z,r)=r-\lambda_{0} z$ and $ f_i^{(1)}(z,r)=-r+\lambda_{0} z$ in (\ref{eq: weak_null_estimand}), we have
 \begin{equation*}
     \theta  = \sum_{i=1}^I\sum_{j=1}^2 \sum_{k=1}^2(-1)^k(r_{ij}^{(k)}-\lambda_{0} z_{i(k)}) = \sum_{i=1}^I\sum_{j=1}^2 \{(r_{ij}^{(2)}-r_{ij}^{(1)})-\lambda_{0} (z_{i(2)}-z_{i(1)})\}.
 \end{equation*}
 Therefore, testing the effect ratio $\lambda = \lambda_0$ is equivalent to testing $\theta = 0$. 
\end{example}

\begin{example}[Thresholded Sample Average Treatment Effect]\label{exp: threshold}
When the treatment is ordinal or continuous, one can consider the sample average difference in potential outcomes for doses above and below some pre-specified threshold $c$. We call this estimand the thresholded sample average treatment effect at threshold c, denoted as TSATE($c$). Formally,
\begin{equation*}
     \text{TSATE}(c) := \frac{1}{N}\sum_{i=1}^I \sum_{j=1}^{n_i}\sum_{k=1}^{n_i}\Big\{ \frac{\mathbbm{1}\{z_{i(k)}>c\}r_{ij}^{(k)}}{m_i} -  \frac{\mathbbm{1}\{z_{i(k)}\leq c\}r_{ij}^{(k)}}{n_i-m_i}\Big\},
 \end{equation*}
 where $m_{i}=\sum_{j=1}^{n_{i}} \mathbbm{1}\{z_{ij} > c\}$ is the number of observed treatment doses in matched set $i$ that are greater than $c$. The TSATE($c$) can be obtained by setting $f_i^{(k)}(z,r) = -(n_i- m_i)^{-1} \times r$ if $z_{i(k)} \leq c$ and $f_i^{(k)}(z,r) = m_i^{-1} \times r$ if $z_{i(k)} > c$ in the general form (\ref{eq: weak_null_estimand}) of the estimand $\theta$. This estimand is similar to an estimand introduced by \citet{Lee2024} in the superpopulation framework.
\end{example}

\begin{example}[Regression-Type Estimands]
We here consider a class of regression-type estimands that can be written in the form of $\theta$ in (\ref{eq: weak_null_estimand}). Specifically, we let $\beta_{ij}$ be the slope coefficient of the simple linear regression of $r_{ij}^{(1)}, \ldots, r_{ij}^{(n_i)}$ on $(z_{i(1)},\ldots, z_{i(n_i)})$: 
\begin{equation*}
 \begin{aligned}
     \beta_{ij} &= \frac{1}{n_i \sum_{k=1}^{n_i} z_{i(k)}^2 - (\sum_{j=1}^{n_i} z_{i(k)})^2}\left(n_i \sum_{k=1}^{n_i} z_{i(k)} r_{ij}^{(k)} - \sum_{k=1}^{n_i} z_{i(k)} \sum_{k=1}^{n_i} r_{ij}^{(k)}\right) \\ 
     &= \frac{1}{\sum_{k=1}^{n_i} z_{i(k)}^2 - \frac{1}{n_i}(\sum_{k=1}^{n_i} z_{i(k)})^2}\left(\sum_{k=1}^{n_i} ( z_{i(k)} - \Bar{z}_{i}) r_{ij}^{(k)}\right),
 \end{aligned}
\end{equation*}
in which $\overline{z}_{i}=n_{i}^{-1}\sum_{j=1}^{n_{i}}z_{ij}$. Then, a regression-type estimand may be some aggregation, such as a weighted average, of these slopes $\beta_{ij}$. For example, one may be interested in the average slope $\overline{\beta}=N^{-1}\sum_{i=1}^{I}\sum_{j=1}^{n_{i}}\beta_{ij}$, which corresponds to taking $f_i^{(k)}(z,r) = (\frac{1}{n_i}\sum_{k=1}^{n_i} z_{i(k)}^2 - \Bar{z}_i^2)^{-1}\times (z-\Bar{z}_i)r / n_i$. 
\end{example}

\begin{example}[Contrast of Stochastic Interventions]\label{exp: stochastic intervention}

In the stochastic intervention setting (\citealp{munoz2012population}; \citealp{Chattopadhyay2023}), assignments of treatment doses to subjects in each matched set follow a prespecified probability distribution. Specifically, suppose that each subject in matched set $i$ had probability $s_{i(k)}$ of receiving dose $z_{i(k)}$, where $\sum_{k=1}^{n_i} s_{i(k)} = 1$ (we denote this distribution as $\mathcal{S}$). Let $\xi^\mathcal{S}_i$ be the expectation of the average outcome in matched set $i$ under the distribution $\mathcal{S}$: $\xi^\mathcal{S}_i = \sum_{k=1}^{n_i}(s_{i(k)}  \frac{\sum_{j=1}^{n_i} r_{ij}^{(k)}}{n_{i}})$. Then, let $\xi^\mathcal{S} = \sum_{i=1}^I \frac{n_i}{N} \xi^\mathcal{S}_i$ be an aggregation of $\xi^\mathcal{S}_i$ from all the matched sets, which corresponds to setting $f_i^{(k)}(z,r) = s_{i(k)} \times r$ in (\ref{eq: weak_null_estimand}). A meaningful estimand is the difference between two such stochastic interventions. For example, the thresholded sample average treatment effect introduced in Example~\ref{exp: threshold} is the difference between the two stochastic interventions, which we will refer to as the above $c$ and below $c$ interventions, respectively. The above $c$ intervention simply puts $s_{i(k)} = 1/m_i$ probability weight on all doses $z_{i(k)} > c$, and $0$ on the rest of doses. The below $c$ intervention simply puts $s_{i(k)} = 1/(n_i-m_i)$ probability weight on all doses $z_i(k) < c$, and $0$ on the rest of doses. Another natural estimand is the difference between an above (or below) $c$ intervention and some baseline intervention (e.g., uniform assignment over the treatment doses in matched set $i$, corresponding to setting $s_{i(k)} = 1/n_i$ for all $z_{i(k)}$ in each matched set $i$). Such estimands for comparing the above c (or below c) and baseline interventions can be respectively expressed as $\xi^{\overline{c},\text{baseline}}=\sum_{i=1}^I \frac{n_i}{N} (\xi_i^{\overline{c}} - \xi_i^{\text{baseline}})$ and $\xi^{\underline{c},\text{baseline}}=\sum_{i=1}^I \frac{n_i}{N} (\xi_i^{\underline{c}} - \xi_i^{\text{baseline}})$, where $\xi_i^{\overline{c}} = \sum_{k=1}^{n_i}\frac{\mathbbm{1}\{z_{i(k)} > c\}}{m_{i}} \frac{\sum_{j=1}^{n_i} r_{ij}^{(k)} }{n_{i}} $, $\xi_i^{\underline{c}} = \sum_{k=1}^{n_i}\frac{\mathbbm{1}\{z_{i(k)} \leq c\}}{n_{i}-m_{i}} \frac{\sum_{j=1}^{n_i} r_{ij}^{(k)} }{n_{i}} $, and $\xi_i^{\text{baseline}} =  \sum_{k=1}^{n_i}\frac{1}{n_i} \frac{\sum_{j=1}^{n_i} r_{ij}^{(k)}}{n_i}$. 
\end{example}

Section~\ref{subsec: analysis for weak nulls} presents methods for estimation, inference, and sensitivity analysis that work for any estimand $\theta$ with the general form (\ref{eq: weak_null_estimand}) following the spirit of Neyman-type weak nulls. 

 \subsection{Estimation, Inference, and Sensitivity Analysis for Neyman-Type Weak Nulls}\label{subsec: analysis for weak nulls}


For the general estimand $\theta$ introduced in (\ref{eq: weak_null_estimand}), we propose the following estimator:
\begin{equation}\label{eqn: estimator for weak nulls}
    V_{N}=\sum_{i=1}^I \frac{n_i}{N} V_{N,i}, \ \text{where $V_{N, i} = \sum_{j=1}^{n_i} \sum_{k=1}^{n_i} f_i^{(k)}(Z_{ij},R_{ij}) \times \mathbbm{1}\{Z_{ij} = z_{i(k)}\}$}.
\end{equation}
\begin{prop}
   Under the no unmeasured confounding assumption, $V_{N}$ is an unbiased estimator for $\theta$, i.e., we have $E[V_{N}\mid\mathcal{F}_{0}, \mathcal{Z}]=\theta$.
\end{prop}
To construct a valid variance estimator for $V_{N}$, we follow a similar procedure to that used for constructing $S_{F}^{2}(Q)$ with the only difference being that we replace each $V_{F,i}$ with $V_{N,i}$. We denote the resulting variance estimator as $S_{N}^{2}(Q)$. Then, as shown in Theorem~\ref{thm: randomization inference for weak null} below, an asymptotically valid sensitivity analysis for testing $H_{N, \theta_{0}}$ defined in (\ref{eq: weak_null_estimand}) can be conducted based on $V_{N}$ and the corresponding variance estimator $S_{N}^{2}(Q)$. 
\begin{theorem}\label{thm: randomization inference for weak null}
 Assuming independence of dose assignments across matched sets and no unmeasured confounding (i.e., the assumption (\ref{eqn: random assignment assumption}) holds), as well as some mild regularity assumptions (see Appendix A), the coverage rate of confidence interval $[V_{N}-\Phi^{-1}(1-\alpha/2)S_N(Q), V_{N}+\Phi^{-1}(1-\alpha/2)S_N(Q)]$ for estimand $\theta$ is asymptotically no less than $100(1-\alpha)\%$.
\end{theorem}

After conducting randomization inference assuming no unmeasured confounding as in Theorem~\ref{thm: randomization inference for weak null}, researchers routinely conduct a sensitivity analysis to assess sensitivity of randomization inference results to unmeasured confounding. To facilitate this, we will extend the results in Theorem~\ref{thm: randomization inference for weak null} to allow existence of unmeasured confounding. Specifically, under the Rosenbaum sensitivity analysis model (\ref{eqn: dose assignment after matching}), given some prespecified sensitivity parameter $\gamma$ (or, equivalently, $\Gamma=\exp(\gamma)$), for each matched set $i$, there exist $ l_{\Gamma,i}$, $h_{\Gamma,i}$ such that $l_{\Gamma,i} \leq p_{i\pi_i} \leq  h_{\Gamma,i}, \ \forall \pi_i\in S_{n_{i}}$, where (in which we define $\mathbf{u}_{i}=(u_{i1},\dots, u_{in_{i}})$)
\begin{equation}\label{eqn: l_i}
    {l}_{\Gamma,i} = \min_{\pi_i, \mathbf{u}_i} \frac{\exp(\gamma(\sum_j z_{i(\pi_i(j))}u_{ij}))}{\sum_{\pi_i'}\exp(\gamma(\sum_j z_{i(\pi_i'(j))}u_{ij}))} =  \min_{\mathbf{u}_i} \frac{\exp(\gamma(\sum_j z_{i(j)}u_{ij}))}{\sum_{\pi_i'}\exp(\gamma(\sum_j z_{i(\pi_i'(j))}u_{ij}))},
\end{equation}
\begin{equation}\label{eqn: h_i}
    {h}_{\Gamma,i} = \max_{\pi_i, \mathbf{u}_{i}} \frac{\exp(\gamma(\sum_j z_{i(\pi_i(j))}u_{ij}))}{\sum_{\pi_i'}\exp(\gamma(\sum_j z_{i(\pi_i'(j))}u_{ij}))} = \max_{ \mathbf{u}_{i}} \frac{\exp(\gamma(\sum_j z_{i(j)}u_{ij}))}{\sum_{\pi_i'}\exp(\gamma(\sum_j z_{i(\pi_i'(j))}u_{ij}))}.
\end{equation}
The final equality in both (\ref{eqn: l_i}) and (\ref{eqn: h_i}) are due to the exchangeability of $u_{ij}$ in the above optimization problems. Although there is no closed-form solution to the optimization problems (\ref{eqn: l_i}) and (\ref{eqn: h_i}), there are many available open-source software packages (e.g., \texttt{R} package \texttt{nloptr}) that are suited for solving such problems efficiently and accurately because the objective functions are smooth and the constraints on the $u_{ij}$ are box constraints. 

Under prespecified sensitivity parameter $\Gamma=\exp(\gamma)$ and prespecified $\theta_{0}$, we define the test statistic for testing $H_{N, \theta_{0}}: \theta=\theta_{0}$ (in which we define $\Gamma_{i}^{*}=h_{\Gamma,i}/l_{\Gamma,i}$): 
\begin{equation*}
   V_{N, \Gamma, \theta_{0}} = \sum_{i=1}^I \frac{n_i}{N} V_{N, \Gamma, \theta_0, i}, \text{where $V_{N, \Gamma, \theta_0, i} = \frac{1+\Gamma_i^*}{2(n_i! h_{\Gamma_i})} \Big( V_{N,i} - \theta_0 - \frac{\Gamma_i^* - 1}{1 + \Gamma_i^*} | V_{N,i} - \theta_0 |\Big)$}.
\end{equation*}
For constructing a variance estimator for $ V_{N, \Gamma}$, we follow a similar procedure to that used for constructing $S^{2}_{F}(Q)$ with the only difference being that we replace each $V_{F,i}$ in $S^{2}_{F}(Q)$ with $V_{N, \Gamma, \theta_0, i}$. We denote the resulting variance estimator as $S^{2}_{N, \Gamma, \theta_{0}}(Q)$, and define the bounding $p$-value $\overline{p}_{N, \Gamma, \theta_{0}}=1-\Phi(V_{N, \Gamma, \theta_{0}} / S_{N, \Gamma, \theta_{0}}(Q))$. Theorem~\ref{thm: V_N weak null validity} shows that $\overline{p}_{N, \Gamma, \theta_{0}}$ is an asymptotically valid $p$-value under $H_{N, \theta_{0}}: \theta=\theta_{0}$, of which the proof involves combining the idea in \citet{Fogarty2023} with some new technical lemmas derived in Appendix A.
\begin{theorem}
\label{thm: V_N weak null validity}
Under the Rosenbaum sensitivity model (\ref{eqn: dose assignment after matching}) and a prespecified sensitivity parameter $\Gamma=\exp(\gamma)$, as well as some mild regularity conditions outlined in Appendix A, we have $\lim_{I \to \infty} P_{\mathbf{u}}[\overline{p}_{N, \Gamma, \theta_{0}}\leq \alpha \mid H_{N, \theta_{0}}, \mathcal{F}, \mathcal{Z}] \leq \alpha$ for any $\alpha\in (0,1)$ and any $\mathbf{u}$.
\end{theorem}

In addition to $V_{N, \Gamma, \theta_{0}}$, we give another class of test statistics $V_{C, \Gamma, \theta_{0}}$ that is also valid for sensitivity analysis for testing Neyman-type weak nulls $H_{N, \theta_{0}}$ if an additional but practically mild condition (as specified in Condition~\ref{condition: V_C sufficient} below) holds. The advantage of $V_{C, \Gamma, \theta_{0}}$ is that it is statistically more powerful than $V_{N, \Gamma, \theta_{0}}$. Specifically, for testing $H_{N, \theta_{0}}$, given sensitivity parameter $\Gamma=\exp(\gamma)$, we define $V_{C, \Gamma, \theta_{0}} = \sum_{i=1}^I \frac{n_i}{N} V_{C, \Gamma,\theta_{0},i }$, where $V_{C, \Gamma,\theta_{0},i}= V_{N, i} - \theta_{0}- \frac{\Gamma_i^p - 1}{1 + \Gamma_i^p} | V_{N, i} - \theta_{0} |$ and $\Gamma_i^p = \max_{\pi_i, \pi_i', \mathbf{u} } p_{i\pi_i} / p_{i\pi_i'} = \exp\Big\{\gamma\Big(\sum_{k = \lceil n_i/2\rceil+1}^{n_i}z_{i(k)}-\sum_{j =1}^{\lfloor n_i/2\rfloor}z_{i(k)}\Big)\Big\}$. For constructing a valid variance estimator for $V_{C, \Gamma, \theta_{0}}$, we follow a similar procedure to that used for constructing $S^{2}_{F}(Q)$ with the only difference being that we replace each $V_{F,i}$ in $S^{2}_{F}(Q)$ with $V_{C, i, \Gamma, \theta_0}$. We denote the resulting variance estimator as $S^{2}_{C, \Gamma, \theta_{0}}(Q)$, and define the bounding $p$-value $\overline{p}_{C,\Gamma, \theta_{0}}=1-\Phi(V_{C, \Gamma, \theta_{0}} / S_{C, \Gamma, \theta_{0}}(Q))$. 

\begin{condition}
\label{condition: V_C sufficient}
Suppose (\ref{eqn: dose assignment in matched set i after matching}) holds and let $\mathbf{u}$ denote the true unmeasured confounders. There exists a vector of unmeasured confounders $\mathbf{u}'$ such that $E_{\mathbf{u}}(V_{C, \Gamma, \theta_{0}} \mid H_{N,\theta_{0}}, \mathcal{F}, \mathcal{Z}) \leq E_{\mathbf{u}'}(V_{C, \Gamma, \theta_{0}} \mid H_{N, \theta_{0} },  \mathcal{F}, \mathcal{Z})$ and $\sum_{i=1}^I \frac{n_i}{N}\frac{\Gamma_i^p-1}{1 + \Gamma_i^p}E_{\mathbf{u}'}(|V_{N,i} - \theta_i| \mid H_{N, \theta_{0} }, \mathcal{F}, \mathcal{Z}) \leq \sum_{i=1}^I \frac{n_i}{N}\frac{\Gamma_i^p-1}{1 + \Gamma_i^p}E_{\mathbf{u}'}(|V_{N,i} - \theta| \mid H_{N, \theta_{0} },  \mathcal{F}, \mathcal{Z})$.
\end{condition}

Roughly speaking, for Condition~\ref{condition: V_C sufficient} to be violated, it would require that on average (weighted by $\frac{n_i}{N}\frac{\Gamma_i^p-1}{1 + \Gamma_i^p}$), $\theta$ is closer to the median of $V_{N,i}$ than $\theta_i$ is. This could happen when there is strong effect heterogeneity and the $\theta_i$ values are very different from $\theta$. Although we expect $V_{C, \Gamma, \theta_{0}}$ to be valid for testing $H_{N,\theta_{0}}$ in a sensitivity analysis for most realistic data generating mechanisms (including those in our simulation settings), in the supplementary material, we give an adversarial example where $V_{C, \Gamma, \theta_{0}}$ is not valid for testing $H_{N, \theta_{0}}$ in a sensitivity analysis. The following theorem verifies the asymptotic validity of using $V_{C, \Gamma, \theta_{0}}$ for sensitivity analysis for testing $H_{N,\theta_{0}}$ under Condition~\ref{condition: V_C sufficient}.
\begin{theorem}
Under the Rosenbaum sensitivity model (\ref{eqn: dose assignment after matching}) and sensitivity parameter $\Gamma=\exp(\gamma)$, Condition \ref{condition: V_C sufficient}, as well as some mild regularity conditions outlined in Appendix A, we have $\lim_{I \to \infty} P_{\mathbf{u}}[\overline{p}_{C, \Gamma, \theta_{0}}\leq \alpha \mid H_{N, \theta_{0}}, \mathcal{F}, \mathcal{Z}] \leq \alpha$ for any $\alpha\in (0,1)$ and any $\mathbf{u}$.
\end{theorem}

\begin{remark}\label{remark: weak null comparison to binary}
When the treatment is binary, our proposed sensitivity analysis methods in Section~\ref{subsec: analysis for weak nulls} reduce to the sensitivity analysis methods for the sample average treatment effect proposed in \citet{Fogarty2023} (see Appendix B for details). 
\end{remark}

\section{Simulation Studies}\label{sec: simulation}

We conduct simulation studies to verify the validity of our methods under Fisher's sharp null and Neyman-type weak nulls, respectively. We first assess the validity of the sensitivity analysis method under Fisher's sharp null $H_{F}$, proposed in Section~\ref{sec: sharp null case}. For each matched set $i$ of the total $I = 1600$ matched sets, we draw $n_i \overset{\text{i.i.d.}}{\sim} \min\{2 + \text{Poisson}(0.6), 4\}$, $z_{i1},\ldots, z_{in_i} \overset{\text{i.i.d.}}{\sim} F_z$, $r_{ij} \overset{\text{i.i.d.}}{\sim} F_{\epsilon}$, and $r_{ij}(z) = r_{ij}$ for all $i, j$ and $z$ (so $H_{F}$ holds). We consider
$F_z \in \{\text{Unif}[0,1], \text{Beta}(2,5),\text{Beta}(2,2)\}$ and $F_{\epsilon}\in \{N(0,1),N(0,5),\text{Exp}(1)-1,\text{Exp}(1/5)-5\}$. Thus, there are 3 choices for the dose distribution $F_z$ and 4 for the baseline outcome distribution $F_{\epsilon}$, leading to $3 \times 4 = 12$ combinations. We use the double-rank test statistic (\citealp{zhang2023statistical}) and set the sensitivity parameter $\Gamma=1.8$ (i.e., $\gamma = \log(1.8)$). Then, over 1000 Monte-Carlo iterations, we draw dose assignments from the randomization distributions according to the ``worst-case" unmeasured confounders that maximize the expectation of the test statistic and conduct hypothesis testing using the proposed bounding $p$-value $\overline{p}_{F}$ based on the test statistic $V_{F}$. All tests were conducted at level $\alpha = 0.1$. According to the simulation results in Table \ref{table: sharp}, all the type-I error rates hover around the $0.1$ significance level under Fisher's sharp null, even under the settings of high-variance baseline outcomes such as $F_{\epsilon}=N(0,5)$ or $\text{Exp}(1/5)-5$. 

\begin{table}[ht]
\caption{Simulation results under Fisher's sharp null case. ``Type-I Error Rate" is the proportion of the 1000 iterations for which the corresponding test using the $p$-value $\overline{p}_{F}$ (based on the bounding test statistic $V_{F}$) falsely rejected the null under level $\alpha = 0.1$. ``Bias," ``SD," and ``Est.SD" denote the mean bias (i.e., the mean difference between the test statistic $T$ employed and its worst-case expectation $\max_{\mathbf{u}}E_{\mathbf{u}}[T\mid H_{F}, \mathcal{F}, \mathcal{Z}]$, of which the value exactly equals the value of $V_{F}$), standard deviation, and estimated standard deviation of the test statistic $V_{F}$ across 1000 simulation runs, respectively. }
\centering
\begin{tabular}{ll|cccc}
  \hline
$F_z$ & $F_\epsilon$ & Type-I Error Rate &  Bias  &  SD  &  Est.SD \\ 
  \hline
Unif[0,1] & N(0,1) & 0.094 & 0.021 & 4.395 & 4.491 \\ 
  Beta(2,5)  & N(0,1) & 0.091 & -0.002 & 4.518 & 4.574 \\ 
  Beta(2,2)  & N(0,1) & 0.106 & 0.102 & 4.738 & 4.683 \\ 
  Unif[0,1] & N(0,5) & 0.101 & -0.135 & 4.468 & 4.458 \\ 
  Beta(2,5)  & N(0,5) & 0.108 & -0.186 & 4.721 & 4.571 \\ 
  Beta(2,2)  & N(0,5) & 0.101 & -0.046 & 4.595 & 4.422 \\ 
  Unif[0,1] & Exp(1/5) & 0.105 & 0.074 & 4.619 & 4.555 \\ 
  Beta(2,5)  & Exp(1/5) & 0.094 & -0.120 & 4.453 & 4.543 \\ 
  Beta(2,2)  & Exp(1/5) & 0.108 & -0.024 & 4.483 & 4.544 \\ 
  Unif[0,1] & Exp(1) & 0.100 & -0.159 & 4.561 & 4.471 \\ 
  Beta(2,5)  & Exp(1) & 0.085 & -0.056 & 4.555 & 4.650 \\ 
  Beta(2,2)  & Exp(1) & 0.086 & 0.001 & 4.245 & 4.425 \\ 
   \hline
\end{tabular}
\label{table: sharp}
\end{table}

Next, we conduct extensive simulation studies to verify the validity of the proposed randomization inference and sensitivity analysis methods based on the bounding test statistics $V_{C, \Gamma, \theta_{0}}$ and $V_{N, \Gamma, \theta_{0}}$ for the Neyman-type weak nulls. For each of the $I = 1000$ matched sets, we draw $n_i \overset{\text{i.i.d.}}{\sim} \min\{2 + \text{Poisson}(1), 5\}$, $z_{i1},\ldots, z_{in_i} \overset{\text{i.i.d.}}{\sim} F_z$, $\beta_i \overset{\text{i.i.d.}}{\sim} F_{\beta}$, $\epsilon_{i(1)k} \overset{\text{i.i.d.}}{\sim} F_{\epsilon_0}$, $\epsilon_{i(j)k} \overset{\text{i.i.d.}}{\sim} F_{\epsilon}$ for $j > 1$, and $r_{ik}(z_{i(j)}) \overset{\text{i.i.d.}}{\sim} \epsilon_{i(j)k} + z_{i(j)} \beta_i$. The dose distributions are chosen from $F_z \in \{\text{Unif}[0,1], \text{Beta}(2,5),\text{Beta}(2,2)\}$. We redraw the doses if there is not at least one dose above and below the threshold $c = 0.5$. The treatment effect distributions $F_{\beta}$, and noise distributions $F_{\epsilon_0}$ and $F_{\epsilon}$ are chosen from $\pm$ Exp(1) $\pm 1$, $\pm$ Exp(1/5) $\pm 5$, $N(0,1)$, and $N(0,5)$, though we do not explore all possible combinations. \text{red}{We also multiply the draws from $F_{\epsilon_0}$ by $\pm B_i$, where $B_i = 2 \times \mathbbm{1}\{\beta_i \geq 0\} - 1$, which induces correlation between the potential outcome under the minimal dose and the direction of the treatment effect.} We consider three levels of sensitivity parameters $\Gamma=\exp(\gamma)$: $\Gamma=1$ (no unmeasured confounding in randomization inference), $\Gamma=1.4$ (moderate unmeasured confounding in sensitivity analysis), and $\Gamma=1.8$ (substantial unmeasured confounding in sensitivity analysis). The estimand is the outcome difference between the above $c$ intervention and the baseline intervention $\xi^{\overline{c},\text{baseline}}$ defined in Example~\ref{exp: stochastic intervention}. We test Neyman's weak null $H_{N, \lambda_{0}}$ under the true value $\lambda_{0}$ of $\xi^{\overline{c},\text{baseline}}$. For both $V_{C, \Gamma, \theta_{0}}$ and $V_{N, \Gamma, \theta_{0}}$, we find the unmeasured confounders that maximize the corresponding expectation (i.e., the worst-case expectation) under $H_{N, \lambda_{0}}$. Then, over 1000 simulation runs, we draw two dose assignments from the randomization distributions according to the worst-case unmeasured confounders for each of $V_{N, \Gamma, \theta_{0}}$ and $V_{C, \Gamma, \theta_{0}}$ and conduct corresponding sensitivity analyses at level $\alpha = 0.1$ (which reduces to randomization inference when there is no unmeasured confounding, i.e., when $\Gamma=1$). We set $Q$ in $S_{C}^{2}(Q)$ and $S_{N}^{2}(Q)$ to only include the intercept term, which is the simplest and most commonly used choice of $Q$. As shown in the simulation results in Table~\ref{table: weak}, all the type-I error rates are below the prespecified significance level $0.1$ and become more conservative as the magnitude of unmeasured confounding $\Gamma$ increases, which agrees with the previous literature in the binary treatment case (\citealp{rosenbaum2020design, Fogarty2023}). Moreover, in all the settings, both $V_{C}$ and $V_{N}$ have mean below zero and valid variance estimates, as guaranteed by the theoretical results in Section~\ref{sec: weak null case}. This also suggests that in a sensitivity analysis (i.e., when $\Gamma>1$), although the worst-case expectations of zero are reachable under particularly adversarial patterns of potential outcomes (as shown in the theoretical results in Section~\ref{sec: weak null case}), they may not be reachable by the allocations of potential outcomes from a regular and more practical data-generating process. One can view this conservatism as a price to be paid for working within the model-free and finite-population sensitivity analysis framework, where almost no assumptions whatsoever are made on the potential outcomes. Moreover, as expected, $V_{C, \Gamma, \theta_{0}}$ is typically less conservative than $V_{N}$. Our results suggest that when testing $H_{N, \lambda_{0}}$, using $V_{C, \Gamma, \theta_{0}}$ instead of $V_{N, \Gamma, \theta_{0}}$ should be more powerful without sacrificing type-I error rate control. Combined with the fact that $V_{C, \Gamma, \theta_{0}}$ is more computationally efficient than $V_{N, \Gamma, \theta_{0}}$ (due to the correction factors $\Gamma_i^p$ in $V_{C, \Gamma, \theta_{0}}$ having a closed form), we recommend using $V_{C, \Gamma, \theta_{0}}$ to test $H_{N, \lambda_{0}}$. This recommendation is similar to a recommendation made by \citet{Fogarty2023} for sensitivity analysis for the sample average treatment effect in the binary treatment case.

\begin{table}[ht]
\caption{Simulation results under the Neyman-type weak nulls case. ``Type-I Error" is the proportion of the 1000 runs for which the corresponding test falsely rejected the null $H_{N,\lambda_{0}}$ under level $\alpha = 0.1$. ``Bias," ``SD," and ``Est.SD" denote the mean bias (i.e., the mean difference between the test statistic employed and its worst-case expectation, of which the value exactly equals the bounding test statistic $V_{C}$ and $V_{N}$), standard deviation, and estimated standard deviation of $V_{C}$ and those of $V_{N}$ across 1000 simulation runs, respectively. }
\footnotesize
\begin{tabular}{llllcc|cccc|cccc}
  \hline
  & & & & & & \multicolumn{4}{c}{$V_{C, \Gamma, \theta_{0}}$} & \multicolumn{4}{c}{$V_{N, \Gamma, \theta_{0}}$} \\ 
 $F_z$ & $F_\beta$ & $F_{\epsilon_0}$ & $F_{\epsilon}$ & $\pm B_i$ & $\Gamma$ & Type-I Error &  Bias & SD & Est.SD & Type-I Error & Bias & SD & Est.SD \\ 
\hline
Beta(2,5) & N(0,1) & N(0,5) & N(0,1) & + & 1.00 & 0.10 & 0.00 & 0.08 & 0.08 & 0.10 & -0.00 & 0.15 & 0.15 \\ 
  Beta(2,5) & N(0,1) & N(0,5) & N(0,1) & + & 1.40 & 0.06 & -0.03 & 0.08 & 0.08 & 0.04 & -0.08 & 0.15 & 0.15 \\ 
  Beta(2,5) & N(0,1) & N(0,5) & N(0,1) & + & 1.80 & 0.03 & -0.05 & 0.08 & 0.07 & 0.01 & -0.14 & 0.15 & 0.15 \\ 
  Beta(2,2) & N(0,1) & N(0,5) & N(0,1) & + & 1.00 & 0.10 & -0.00 & 0.10 & 0.10 & 0.10 & -0.00 & 0.20 & 0.20 \\ 
  Beta(2,2) & N(0,1) & N(0,5) & N(0,1) & + & 1.40 & 0.04 & -0.05 & 0.11 & 0.11 & 0.04 & -0.13 & 0.22 & 0.21 \\ 
  Beta(2,2) & N(0,1) & N(0,5) & N(0,1) & + & 1.80 & 0.01 & -0.08 & 0.10 & 0.11 & 0.01 & -0.23 & 0.22 & 0.22 \\ 
  Beta(2,5) & N(0,1) & Exp(1/5) & N(0,1) & + & 1.00 & 0.10 & 0.00 & 0.08 & 0.08 & 0.10 & 0.00 & 0.15 & 0.15 \\ 
  Beta(2,5) & N(0,1) & Exp(1/5) & N(0,1) & + & 1.40 & 0.06 & -0.02 & 0.08 & 0.08 & 0.04 & -0.07 & 0.16 & 0.16 \\ 
  Beta(2,5) & N(0,1) & Exp(1/5) & N(0,1) & + & 1.80 & 0.03 & -0.05 & 0.07 & 0.07 & 0.02 & -0.13 & 0.15 & 0.15 \\ 
  Beta(2,2) & N(0,1) & Exp(1/5) & N(0,1) & + & 1.00 & 0.10 & 0.00 & 0.10 & 0.10 & 0.09 & -0.00 & 0.20 & 0.21 \\ 
  Beta(2,2) & N(0,1) & Exp(1/5) & N(0,1) & + & 1.40 & 0.04 & -0.05 & 0.10 & 0.10 & 0.03 & -0.12 & 0.20 & 0.20 \\ 
  Beta(2,2) & N(0,1) & Exp(1/5) & N(0,1) & + & 1.80 & 0.02 & -0.08 & 0.10 & 0.10 & 0.01 & -0.22 & 0.21 & 0.20 \\ 
  Beta(2,5) & N(0,1) & -Exp(1/5) & N(0,1) & + & 1.00 & 0.11 & 0.00 & 0.07 & 0.07 & 0.10 & -0.00 & 0.15 & 0.15 \\ 
  Beta(2,5) & N(0,1) & -Exp(1/5) & N(0,1) & + & 1.40 & 0.06 & -0.02 & 0.08 & 0.08 & 0.04 & -0.07 & 0.16 & 0.16 \\ 
  Beta(2,5) & N(0,1) & -Exp(1/5) & N(0,1) & + & 1.80 & 0.03 & -0.05 & 0.08 & 0.08 & 0.02 & -0.13 & 0.16 & 0.16 \\ 
  Beta(2,2) & N(0,1) & -Exp(1/5) & N(0,1) & + & 1.00 & 0.12 & 0.00 & 0.10 & 0.11 & 0.10 & 0.00 & 0.20 & 0.21 \\ 
  Beta(2,2) & N(0,1) & -Exp(1/5) & N(0,1) & + & 1.40 & 0.04 & -0.04 & 0.10 & 0.10 & 0.04 & -0.12 & 0.20 & 0.21 \\ 
  Beta(2,2) & N(0,1) & -Exp(1/5) & N(0,1) & + & 1.80 & 0.02 & -0.09 & 0.11 & 0.10 & 0.02 & -0.22 & 0.22 & 0.21 \\ 
  Beta(2,5) & N(0,1) & N(0,5) & N(0,1) & $-$ & 1.00 & 0.08 & -0.00 & 0.08 & 0.08 & 0.11 & 0.00 & 0.16 & 0.16 \\ 
  Beta(2,5) & N(0,1) & N(0,5) & N(0,1) & $-$ & 1.40 & 0.04 & -0.03 & 0.08 & 0.08 & 0.04 & -0.08 & 0.15 & 0.15 \\ 
  Beta(2,5) & N(0,1) & N(0,5) & N(0,1) & $-$ & 1.80 & 0.03 & -0.04 & 0.08 & 0.08 & 0.01 & -0.14 & 0.15 & 0.16 \\ 
  Beta(2,2) & N(0,1) & N(0,5) & N(0,1) & $-$ & 1.00 & 0.10 & 0.00 & 0.11 & 0.10 & 0.09 & 0.00 & 0.21 & 0.21 \\ 
  Beta(2,2) & N(0,1) & N(0,5) & N(0,1) & $-$ & 1.40 & 0.04 & -0.05 & 0.10 & 0.10 & 0.04 & -0.13 & 0.21 & 0.21 \\ 
  Beta(2,2) & N(0,1) & N(0,5) & N(0,1) & $-$ & 1.80 & 0.02 & -0.08 & 0.10 & 0.10 & 0.01 & -0.23 & 0.21 & 0.21 \\ 
  Beta(2,5) & N(0,1) & Exp(1/5) & N(0,1) & $-$ & 1.00 & 0.09 & -0.00 & 0.07 & 0.07 & 0.10 & -0.00 & 0.14 & 0.14 \\ 
  Beta(2,5) & N(0,1) & Exp(1/5) & N(0,1) & $-$ & 1.40 & 0.05 & -0.03 & 0.07 & 0.07 & 0.02 & -0.08 & 0.15 & 0.15 \\ 
  Beta(2,5) & N(0,1) & Exp(1/5) & N(0,1) & $-$ & 1.80 & 0.02 & -0.04 & 0.07 & 0.07 & 0.02 & -0.13 & 0.15 & 0.15 \\ 
  Beta(2,2) & N(0,1) & Exp(1/5) & N(0,1) & $-$ & 1.00 & 0.10 & 0.00 & 0.11 & 0.11 & 0.11 & 0.02 & 0.22 & 0.22 \\ 
  Beta(2,2) & N(0,1) & Exp(1/5) & N(0,1) & $-$ & 1.40 & 0.05 & -0.05 & 0.11 & 0.11 & 0.04 & -0.11 & 0.21 & 0.22 \\ 
  Beta(2,2) & N(0,1) & Exp(1/5) & N(0,1) & $-$ & 1.80 & 0.02 & -0.09 & 0.11 & 0.11 & 0.01 & -0.22 & 0.21 & 0.22 \\ 
  Beta(2,5) & N(0,1) & -Exp(1/5) & N(0,1) & $-$ & 1.00 & 0.10 & -0.00 & 0.07 & 0.07 & 0.11 & 0.00 & 0.15 & 0.15 \\ 
  Beta(2,5) & N(0,1) & -Exp(1/5) & N(0,1) & $-$ & 1.40 & 0.04 & -0.03 & 0.07 & 0.08 & 0.04 & -0.07 & 0.15 & 0.15 \\ 
  Beta(2,5) & N(0,1) & -Exp(1/5) & N(0,1) & $-$ & 1.80 & 0.03 & -0.05 & 0.07 & 0.07 & 0.02 & -0.14 & 0.15 & 0.15 \\ 
  Beta(2,2) & N(0,1) & -Exp(1/5) & N(0,1) & $-$ & 1.00 & 0.11 & 0.01 & 0.10 & 0.10 & 0.09 & 0.01 & 0.19 & 0.19 \\ 
  Beta(2,2) & N(0,1) & -Exp(1/5) & N(0,1) & $-$ & 1.40 & 0.04 & -0.04 & 0.10 & 0.11 & 0.03 & -0.13 & 0.21 & 0.21 \\ 
  Beta(2,2) & N(0,1) & -Exp(1/5) & N(0,1) & $-$ & 1.80 & 0.02 & -0.09 & 0.10 & 0.10 & 0.01 & -0.23 & 0.21 & 0.21 \\ 
   \hline
\end{tabular}
\label{table: weak}

\end{table}

\section{Data Application}

We apply our methods to reanalyze a study on the effect of lead treatment on lumbar bone mineral density (BMD) among young female adults (\citealp{Lu2021}). The data was obtained from the records of females aged 20-39 in the 2011--18 National Health and Nutrition Examination Survey (NHANES). The continuous treatment variable is the log of the lead concentration in the blood measured in micrograms per deciliter. The outcome variable is lumbar spine BMD measured in grams per centimeter squared. We controlled for the same set of covariates as in \citep{Lu2021}: age, weight, height, race, family income poverty ratio, smoking status, physical activity level, albumin level, blood urea nitrogen level, uric acid level, phosphorus level, and calcium level, as well as missingness indicators for these covariates. The estimand of interest is the difference between outcome under the below $c$ stochastic intervention and the baseline intervention $\xi^{\underline{c},\text{baseline}}$ defined in Example~\ref{exp: stochastic intervention}, with threshold $c$ chosen to be 0.74675 $\mu g / dL$, which is the mean of the 4 geometric mean blood levels among females above age 1 in NHANES waves 2011-12, 2013-14, 2015-16, and 2017-18. We adopt the commonly used nonbipartite matching algorithm of \citet{zhang2023statistical} based on Mahalanobis distance. We discarded matched sets with pairwise mean Mahalanobis distance exceeding 7 (roughly 25\% of the samples), which leaves us 1436 subjects in 711 matched sets. To ensure sufficient covariate balance after matching, we tested the post-matching randomization assumption (\ref{eqn: random assignment assumption}) using a variant of the classification permutation test \citep{Gagnon-Bartsch2019, Chen2023}, proposed in the \citet{zhang2024sensitivity}, and fail to reject at significance level 0.1. This indicates that the observed covariates (measured confounders) are sufficiently balanced after matching. We conduct one-sided randomization tests for Fisher's sharp null of no effect $H_{F}$ (based on $V_{F}$ using double-rank test statistic) and construct 95\% confidence intervals (based on $V_{C}$) for $\xi^{\underline{c},\text{baseline}}$, both under $\Gamma=1$ (i.e., randomization inference) and $\Gamma>1$ (sensitivity analysis). From the results displayed in Table \ref{tab:lead_bmd_results}, we see that there is some evidence against Fisher's sharp null of no effect $H_{F}$, with the $p$-value crossing the conventional $\alpha = 0.05$ level at $\Gamma = 1.30$. Meanwhile, the confidence intervals suggest that shifting the population's log lead exposure below $c$ could have a positive effect on bone mineral density. For reference, the standard deviation of the outcome in the matched sample was roughly 0.12. The lower bound of the 95\% confidence interval crosses zero around $\Gamma = 1.20$.

\begin{table}[ht] 
    \caption{Randomization inference and sensitivity analysis results of the real data application. The point estimate (under the no unmeasured confounding assumption, i.e., under $\Gamma=1$) for $\xi^{\underline{c},\text{baseline}}$ is 0.0128.}
    \centering
    \begin{tabular}{l|rr}
        $\Gamma$ & $p$-value under $H_{F}$ &  90\% CI for $\xi^{\underline{c},\text{baseline}}$
        \\ \hline
        1.00 & 0.0003 &   [0.0050, 0.0207]  \\  
        1.05 & 0.0012 &  [0.0035, 0.0222]  \\  
        1.10 & 0.0039 & [0.0021, 0.0236]  \\  
        1.15 & 0.0106 & [0.0008, 0.02496]  \\  
        1.20 & 0.0251 & [-0.0005, 0.0263] \\  
        1.25 & 0.0519 &  [-0.0018, 0.0275]\\  
        1.30 & 0.0951 & [-0.0029, 0.0287] \\
        \hline
    \end{tabular}
    \label{tab:lead_bmd_results}
\end{table}

\section{Discussion}

In this work, we have developed novel randomization inference and sensitivity analysis methods for matched observational studies with treatment doses (e.g., continuous or ordinal treatments). Our methods apply to general matching (and stratification) designs with treatment doses, general outcome variables (e.g., binary or non-binary), and cover both Fisher's sharp null and Neyman-type weak nulls (the two major types of null hypotheses in causal inference). Therefore, our methods fill several important gaps between the abundance of matching designs with treatment doses and the lack of universally applicable randomization inference and/or sensitivity analysis methods for matching with treatment doses.




\section*{Acknowledgements}

The authors thank Professors Dylan Small and Eric Tchetgen Tchetgen for helpful advice and comments. The work of Siyu Heng is in part supported by NIH grant R21DA060433.


\section*{Supplementary Materials}

The online supplementary materials include technical proofs and additional discussions, remarks, and illustrative examples. 








\bibliographystyle{apalike}
\bibliography{references}

\end{document}